\documentclass[final,5p,sort&compress]{elsarticle}

\usepackage{lineno,hyperref}
\modulolinenumbers[5]

\journal{arXive}

\usepackage{lineno,hyperref}
\modulolinenumbers[5]
\usepackage[latin1]{inputenc}
\usepackage{amsmath,empheq}
\usepackage{amsfonts}
\usepackage{amssymb}
\usepackage[only,llbracket,rrbracket]{stmaryrd}
\usepackage{graphicx}
\usepackage{adjustbox}
\graphicspath{{./figures/}}
\usepackage[font=small,format=plain,labelfont=bf,up,textfont=normal,up,justification=justified,singlelinecheck=false]{caption}
\usepackage{subfig}
\usepackage{float}
\usepackage{color}
\usepackage{booktabs}
\usepackage[normalem]{ulem}

\usepackage{tabularx}

\newcolumntype{L}[1]{>{\raggedright\arraybackslash}p{#1}}
\newcolumntype{C}[1]{>{\centering\arraybackslash}p{#1}}
\newcolumntype{R}[1]{>{\raggedleft\arraybackslash}p{#1}}

\begin{document}
\begin{frontmatter}

\title{Modeling of internal mechanical failure of all-solid-state batteries during electrochemical cycling, and implications for battery design}

\author[mymainaddress]{Giovanna Bucci
\corref{mycorrespondingauthor}}
\cortext[mycorrespondingauthor]{Corresponding author}
\ead{bucci@mit.edu}

\author[mymainaddress]{Tushar Swamy}
\author[mymainaddress]{Yet-Ming Chiang}
\author[mymainaddress]{W. Craig Carter}

\address[mymainaddress]{Massachusetts Institute of Technology, Department of Materials Science and Engineering -
77 Massachusetts Avenue, Cambridge, MA 02139-4307 USA}

\begin{abstract}
This is the first quantitative analysis of
%study to explore 
mechanical reliability of all-solid state batteries.
Mechanical degradation of the solid electrolyte (SE) is caused by intercalation-induced expansion of the electrode particles, within the constrains of a dense microstructure.
A coupled electro-chemo-mechanical model was implemented to quantify the material properties that cause a SE to fracture.
The treatment of microstructural details 
is essential to the understanding of stress-localization phenomena and fracture.
A cohesive zone model is employed to simulate the evolution of damage.
In the numerical tests, fracture is prevented when
%only if 
electrode-particle's expansion is lower than 7.5\% 
(typical for most Li-intercalating compounds)
and the solid-electrolyte's fracture energy higher than $G_c = 4$ J m$^{-2}$. 
Perhaps counter-intuitively, the analyses show that 
compliant solid electrolytes (with Young's modulus in the order of E$_{SE} = 15$ GPa) are more prone to micro-cracking. 
This result, 
captured by our non-linear kinematics model, 
contradicts
the speculation that sulfide SEs are more suitable
for the design of bulk-type batteries than oxide SEs.
Mechanical degradation is linked to the battery power-density. Fracture in solid Li-ion conductors represents a barrier for Li transport, and accelerates the decay of rate performance.
\end{abstract}

\end{frontmatter}

%\linenumbers

%\gioshort{Last update on Feb 3 at 11.02am by Gio}

\section{Introduction}

Li-ion batteries that use solid electrolyte materials (SEs) in place of traditional liquid electrolytes could achieve high energy density while avoiding safety issues surrounding liquid electrolyte flammability.
Solid electrolytes with conductivity approaching that of liquid electrolyte have recently been discovered~\cite{Takada:2013,Li2015, Kim2015, ASSBgarnet_rev2014, C0CS00081G, Takada2014, Tatsumisago2013}. 
Despite fast growing interest in all-solid-state batteries (ASSBs), many challenges remain in both manufacturing and reliability of the technology. 

In ASSBs, the solid-electrolyte is responsible for binding the active material and establishing conductive paths for Li ions. 
However, the formation of micro-cracks within the solid electrolyte is expected to reduce its effective ionic conductivity.
Additionally,
low porosity solid-state systems are expected to be more prone to mechanical degradation if not designed to accommodate intercalation-induced deformations.
As fracture degrades the microstructure, paths for lithium diffusion become more tortuous and the battery power-density decreases.
Microscale defects and inhomogeneities in battery microstructures would  interfere with Li transport and accelerate the decay of battery performance.
%Furthermore, micro-cracks may allow for Li dendrites to penetrate dense electrodes, causing the cell to short~\cite{Suzuki2015172}.
Furthermore, at sufficiently high current densities, micro-cracks may provide a pathway for Li dendrite growth, eventually causing the cell to short~\cite{Suzuki2015172}.

We employ a fully coupled electro-chemo-mechanical model to investigate fracture mechanisms
in composite solid-state electrodes. 
Treatment of microstructural details and local variability~\cite{bucciActaMat2015} enables the study of stress localization caused by
particle misalignment and non-smooth features. 
A cohesive zone model (CZM) is employed to simulate the evolution of damage~\cite{bucci2016book}. The detrimental effect of fracture on Li-ion flux is also taken into account by the CZM.
We quantify the conditions under which fracture occurs, caused by the chemical expansion of electrode particles.
Fracture is the result of 
regions of shear and tensile stresses formed during electrochemical cycling.
%are predicted to form and eventually cause fracture.

\begin{table*}[]
\centering
\caption{Mechanical properties of solid electrolyte materials.}
\label{tab:SEmechanicalProp}
\begin{adjustbox}{width=1\textwidth}
\begin{tabular}{ l  L{3.5cm}  L{3cm}  L{3cm}  L{2cm}  L{3cm}  L{1.5cm}  L{3cm} }
\hline
& Compound	&	Processing Method	& Young's Modulus 	 & Fracture Toughness 	& Testing Method 	& Reference  & Conductivity  \newline (room temperature)   \\
\hline
LIPON  & 
Li$_x$PO$_y$N$_z$  & 
{\small{Amorphous LiPON films magnetron sputtered}} & 
$77$ GPa  & 
& 
nanoindentation    & 
\cite{Herbert2011} & 
2 10$^{-6}$ S cm$^{-1}$  \\
\midrule
Perovskite & 
Li$_{0.33}$La$_{0.57}$TiO$_3$ - solid state  & 
{\small{Hot-pressing at $1000^{\circ}$ C, \newline
relative density \textgreater 95\%}} & 
$186 \pm 4 $ & 
$0.890 - 1.34$ MPa m$^{0.5}$ & 
nanoindentation    & 
\cite{Cho2012} & 
bulk: $\sim$ 10$^{-3}$ S cm$^{-1}$ \\
 & 
Li$_{0.33}$La$_{0.57}$TiO$_3$ - sol gel&   
 & 
$200 \pm 3$ GPa      & 
$0.890 - 1.31$ MPa m$^{0.5}$ &                                                      nanoindentation&    
&  
total: $\sim$ 10$^{-5}$ S cm$^{-1}$   \\
\midrule
Garnet & 
Li$_{6.24}$La$_3$Zr$_2$Al$_{0.24}$O$_{11.98}$ (LLZO)  & 
{\small{Hot pressed}}  & 
150 GPa (porosity=0.03);     $132.5$ GPa \newline (porosity = 0.06)&  
& 
resonant ultrasound spectroscopy    & 
\cite{Ni2012} & 
$\sim$ 0.2 10$^{-3}$ S cm$^{-1}$ \\
& 
cubic Li$_7$La$_3$Zr$_2$O$_{12}$               
& {\small{Relative density of $\sim$9~ \%}} & 
150 GPa &                       
& 
resonant ultrasound spectroscopy  & 
& 
3 0$^{-4}$ S cm$^{-1}$ \\
\midrule
%NASICON type glass-ceramic & Li$1+x+y$Al$x$Ti$2-x$Si$y$P$3-y$O$12$ (LATP) & relative density of $\sim$100 \%                                & 120 GPa                     &                       & nanoindenation                                       & \cite{}                & $\sim$ 1.2 10$^{-3}$ S cm$^{-1}$ \\
%                           & (x and y in 0.1-0.3 range)       &                                                                 &                             &                       &                                                      &                                               &                                       \\
%\midurle
Sulfide & 
Li$_2$S-P$_2$S$_5$  - hot pressed         & 
{\small{Sintering at 360 MPa and temperatures $20-190^{\circ}$ C        }} & 
$18-25$ GPa                   &                       
& 
ultrasound velocity measurement and compression test & 
\cite{Sakuda2013youngMod} & 
 3 10$^{-4}$ S cm$^{-1}$ 
(for fully dense material) \\
& 
Li$_2$S-P$_2$S$_5$ - cold pressed        & 
{\small{Sintering at 180-360 MPa and room temperature}} & 
14-17 GPa                   & 
&  
&                   &   \\
 & 
Li$_2$S-P$_2$S$_5$         & 
{\small{}} & 
$18.5 \pm 0.9$ GPa                   &     
$0.23 \pm 0.04$ MPa m$^{0.5}$                  
& 
nanoindentation & 
\cite{McGrogan2016} & 
  \\
 & 
Li$_{10}$GeP$_2$S$_{12}$ &                                                                 & 
$37.19$ GPa    &                       
 & 
Atomistic simulation   & 
\cite{Wang2014} & 
1.2 10$^{-2}$ S cm$^{-1}$ \\
%                           &                                  &                                                                 & (seemingly ductile)         &                       &                                                      &   
\hline                                           &                                      
\end{tabular}
\end{adjustbox}
\end{table*}

The role of intercalation-induced stress (also called Vegard's stress) on the mechanical failure of electrode particles has been previously studied~\cite{woodford2010electrochemical,Bhandakkar20101424, BowerGuduru:2012, Renganathan01022010, Garcia01012005, Wang01112007, Golmon20091567, Zhu01012012, Purkayastha2012, NME:NME5133, Aifantis2005203, Ryu20111717, Aifantis20112122, Kalnaus20118116, Xia201478, Zhang201547,  Hao2013415, Chew20144176, Drozdov201467, Ye2014447, Greve2012377, Wang2012236, Pharr2014569, Min2015835, Zhang2015102, Ma2015114,Zhang2015309, Damle2016373, Laresgoiti, Ryu2014274,Dai, Yang, Dimitrijevic, Lee201637}.
Among these studies,
only Bower and Guduru~\cite{BowerGuduru:2012}
employed a fully coupled chemo-mechanical model to
%a few of these studies 
simulate fracture of a simplified electrode microstructure.
To our knowledge, 
ours is the first model to
quantitatively assess mechanical reliability of all-solid state batteries, and predict the extension of fracture caused by electrochemical cycling.

The mechanical properties of solid electrolyte materials have not received much attention and very limited experimental chemo-mechanical properties are available. 
Measurements collected in Table~\ref{tab:SEmechanicalProp} reveal a wide range of values for Young's modulus. In particular, sulfide SEs tend to be much more compliant than oxide electrolytes.
The Young's modulus of Li$_2$S-P$_2$S$_5$ sulfide solid electrolytes 
has been estimated to be in the range of 
14-25 GPa~\cite{Sakuda2013youngMod, McGrogan2016}

Such a low stiffness
has been regarded as favorable
for the design of bulk-type batteries~\cite{Sakuda2013}.
However, we show that compliant solid electrolytes (with Young's moduli in the order of E$_{SE} = 15$ GPa) are more prone to micro-cracking. Solid electrolytes deform by stretching and shearing in response to the particles' volume change.
The nonlinear formulation of the mechanical equilibrium
quantifies the 
difference in deformation and stress patterns 
associated with varying the SE's stiffness.  
A linear model, which would predict that stress scales with Young's modulus, 
would not capture the the microstructural effects that we describe below.

We compare the evolution of damage for several values of electrolyte's fracture energy ($G_c = 0.25 - 4.0$ J m$^{-2}$) and volume expansion of the active material (7.5\%, 15\% and 30\%).
A cohesive model postulates that fracture energy is released gradually as the crack opens. The CZM differs from the Griffith model wherein energy is released instantaneously.
%
%Cohesive energy models are based on the premise that the fracture energy, instead of being
%released instantaneously at crack initiation as in Griffith's model, is released gradually with
%the growth of the crack opening. 
The gradual release presumes some cohesion between
the separating flanks of a crack.
Generally, the traction decays with increasing separation until it vanishes at a critical opening displacement. 
The fracture energy represents the integral of the traction-separation curve and it is treated as a model-parameter.
In our analyses, cracking is prevented only 
in those cases for which
the electrolyte's fracture energy $G_c \geq 4.0$ J m$^{-2}$ and the particles' total volumetric expansion is $\Delta V \leq 7.5$\%. In all the other cases, the model predicts some extension of mechanical degradation.

Recent studies have analyzed the properties of the interface between solid electrolytes and electrode materials~\cite{C5TA08574H,RichardsCeder,Wenzel201624}.
In particular, Zhu \emph{et al.}~\cite{C5TA08574H} have identified voltage-stability ranges for many SE materials and predicted decomposition products of the interface reaction.
The formation of an interface layer is expected to affect lithium transfer kinetics. Thick interface layers may also modify the local mechanical response. 
However, further studies are needed to characterize the interfacial properties. We believe the assumption of a perfectly coherent and stable interface to be appropriate for the scope of this study.

In the following section we illustrate the model and discuss the results in detail.

%%%%%%%%%%%%%%%%%%%%%%%%%%%%%%%

\section{Modeling of fracture in all-solid-state battery electrodes}
\label{sec:fracture_SE}

\subsection*{Methods}

\begin{figure*}
\centering
\includegraphics[width=0.7\textwidth]{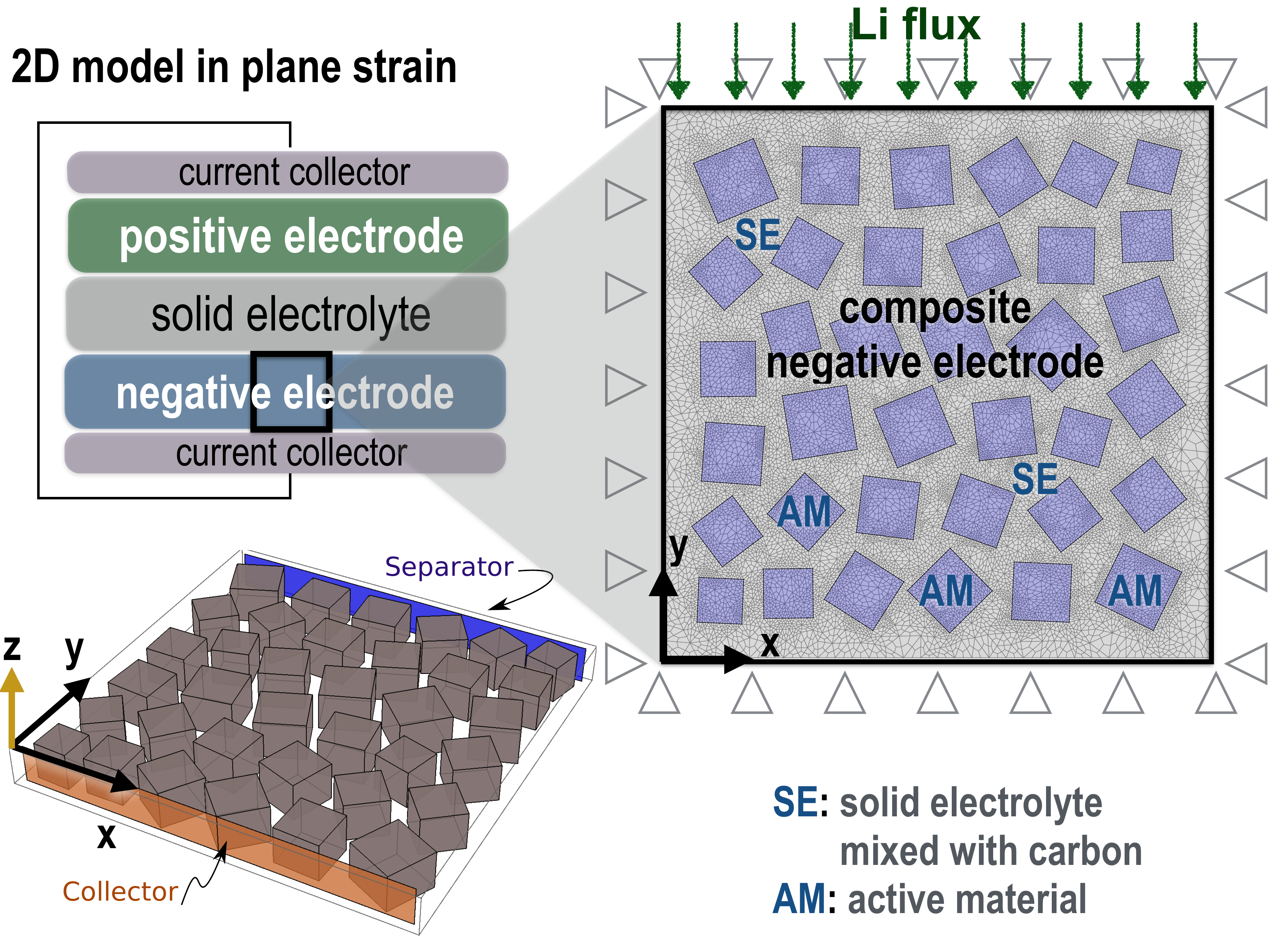} 
\caption{Geometry, discretization and boundary conditions of a finite element model of a composite electrode. Electrode particles are  embedded in a mixed conductor, consisting of a solid electrolyte and an electronically conductive additive.}
\label{Fig3:mesh}
\end{figure*}
In one common design of ASSBs, the positive and negative electrodes are composites of active electrode-particles embedded in a solid-matrix admixture of ionic and electronic conducting materials.
Negative electrodes are, in most cases, produced and assembled in the delithiated state. During the first charge, the anode particles tend to expand as they intercalate Li 
(experimentally measured Vegard's strains for many Li-storage compounds are summarized in Tables 1.1 and 1.2 of Woodford's thesis.~\cite{woodford:thesis}, and in Table 1 of Mukhopadhyay \& Sheldon~\cite{Mukhopadhyay2014}).
Constrained by the surrounding SE matrix, the particles will be in compression.
The  mechanical stress and degradation caused by swelling of electrode particles is modeled as follows.
%We study here mechanical stress and degradation caused by swelling of electrode particles.

A finite element (FE) code was implemented according to the theoretical continuum model described by Bucci \emph{et al.}~\cite{bucciActaMat2015}. 
A representative microstructural arrangement of a distribution of particle sizes
and its FE discretization with linear quadrilateral elements are represented in Fig.~\ref{Fig3:mesh}. 
The FE mesh is representative of a portion of the composite negative electrode (square highlighted in Fig.~\ref{Fig3:mesh}). The section extends from the current collector to the separator. The direction parallel to the Li flux is marked with \emph{y} in Fig.~\ref{Fig3:mesh}. 
%The dimension is bounded to the power density of the cell.
Because of the large computational cost of a full 3D analysis, the system is modeled in 2D under the assumption of plane strain. 
According to the 2D plane strain model, the electrode particles are allowed to expand in the \emph{xy} plane , but not in the \emph{z} direction.
Plane strain is typically employed for a thin plate embedded in a thicker sample.
This is a realistic assumption for a typical bulk-type electrode. The  2D model is expected  to correctly capture trends in stress and fracture.

The grid is managed with the deal.II finite-element library~\cite{BangerthHartmannKanschat2007,dealII82}.
The microstructure includes 36 randomly oriented square particles in a region of dimensions $ 11 \mu m \; \times \; 11 \mu m$. 
The particle's position and size distribution follow from a centroidal Voronoi tessellation.
The average particle size is $ 1 \mu m $. The area ratio of active material is about 50\%--a typical volume ratio for commercial Li-ion batteries is about 50-60\%.
We consider shapes with sharp corners (the squares chosen here) to be a more realistic representation of particles than circles--see for instance Fig.~6 of Sakuda \emph{et al.}~\cite{Sakuda2013}
and 3D image reconstruction of LiCoO$_2$ particles of Harris \emph{et al.}~\cite{Harris2013}.
Flaws and stresses are more likely to accumulate near sharp corners.
% and therefore corners are likely to be associated with damage.

At a given time step, the electrochemical-mechanical problem is solved employing a Newton-Raphson iterative algorithm. At each iteration, 
displacements, Li concentration and diffusion potential
are calculated, as the solution to 
%three coupled equations.
three equations that couple the electro-, chemo-, and mechanical fields.
We model the matrix as a homogenized admixture of ionic and electronic conductors. 
%\gioshort{Lithium-ion transport is charge-compensated by electrons diffusing through the electronically conductive material embedded in the matrix itself.}
The constitutive behavior for the electrode and the electrolyte material is assumed to be elastically and diffusively isotropic. 
Materials are assumed to have a linear elastic constitutive behavior.
%This investigation is limited to the linear elastic regime. 
The solid electrolyte material is considered to have zero Vegard's strain. 
Input parameters for this problem  are summarized in Table.~\ref{tab:fracture_SE} (the variables listed appear in the equations discussed in Bucci \emph{et al.}~\cite{bucciActaMat2015}).

The CZM of fracture is based on an intrinsic history-dependent constitutive behavior~\cite{bucci2016book}.
The flux across the interface is irreversibly set to zero at the onset of fracture.
%(i.e., once the opening displacement reaches the value corresponding to the elastic limit.
%Following \cite{ortizpandolfi:1999, bower:2012}, we assume the bilinear traction-separation law.
%represented in Fig~.\ref{CZM_tsl}. 
%
%
The "intrinsic" CZM approach is based on the pre-insertion of cohesive elements along potential crack paths.
Therefore, fracture is allowed to propagate along a subset of finite element interfaces. Those interfaces lie along crack patterns that run between particles.
% corners, where fracture is most likely to develop. 
The pre-insertion of cohesive elements along potential crack paths restrict the propagation of fracture.
%follows from the "intrinsic" approach chosen here and discussed in Ref.~\ref{sec:CZM_constitutive}.
An alternative approach, called "extrinsic", is based on the insertion of cohesive elements on the fly, only at the interfaces where the fracture criterion is met. However, this method requires changes in the mesh topology, and it is not suitable for parallel computing. 
%The communication among processors becomes more and more expensive as fracture extends.
We assume that the placement of the CZ elements in regions of high shear would not differ from the extrinsic approach.

%; see Fig.~\ref{CZM_tsl}). 
%{\color{red}{Further details on the cohesive zone model can be found in~\ref{sec:CZM_constitutive}.}}
%

Galvanostatic tests are performed by applying a constant and uniform  lithium flux (corresponding to a constant current density) at the separator interface (top edge in Fig.~\ref{Fig3:mesh}). 
The spatial variability of the current density at the separator depends on specific features of the electrochemical cell, such as electrode thickness and tortuosity.
%For this study, we will assume constant current density.
%The uniform current density is an approximation. 
Polarization effects in proximity of the separator could be modeled by treating both sides of the electrochemical cell.
A zero flux is assumed on the remaining edges.

For the mechanical problem, zero horizontal displacement Dirichlet boundary conditions are applied on the left and right boundaries.
The displacement is considered fully constrained (in the vertical direction) on the top edge by the presence of the SE, and at the bottom edge by packaging or neighbor cells. 
We expect the volume change of the active material to be accommodated by the deformation of the SE matrix in ASSBs.
This is similar to what Harris \emph{et al.}~\cite{Harris2013} observed (by
digital image correlation techniques) in graphite electrodes, where most
of graphite's swelling was compensated by reducing the electrode porosity.
Harris \emph{et al.}~showed that the average strain of a graphite electrode was only about 0.2\% during lithiation, an order of magnitude smaller than graphite's chemical expansion. 

\subsection*{Results}

For the baseline example, we choose a   
solid electrolyte material, having
Young's modulus E$_{SE} = 15$~GPa 
and bulk fracture energy $G_c = 1.0$ J m$^{-2}$.
These are representative values a sulfide SE material.
%
%and an intercalating compound with 30\% volumetric expansion at full lithiation.
%($ \beta_{Li} = 0.1 $).
%Our simulation shows that fracture initiates when particles have changed their volume by only 3\%, corresponding to 10\% of the total lithiation.
In order to reveal behavior over a wide range of Vegard's expansion, the intercalation compound was allowed to have up to 30\% volumetric expansion at full lithiation.  However, our simulation shows that fracture initiates when particles have changed their volume by only 3\%, a value that encompasses the behavior of many intercalation compounds.

\begin{figure} 
\centering
\subfloat[t = 150 s, SOC = 0.115641] {
\label{fig:fractureSE_1}
\includegraphics[width=0.45\textwidth]{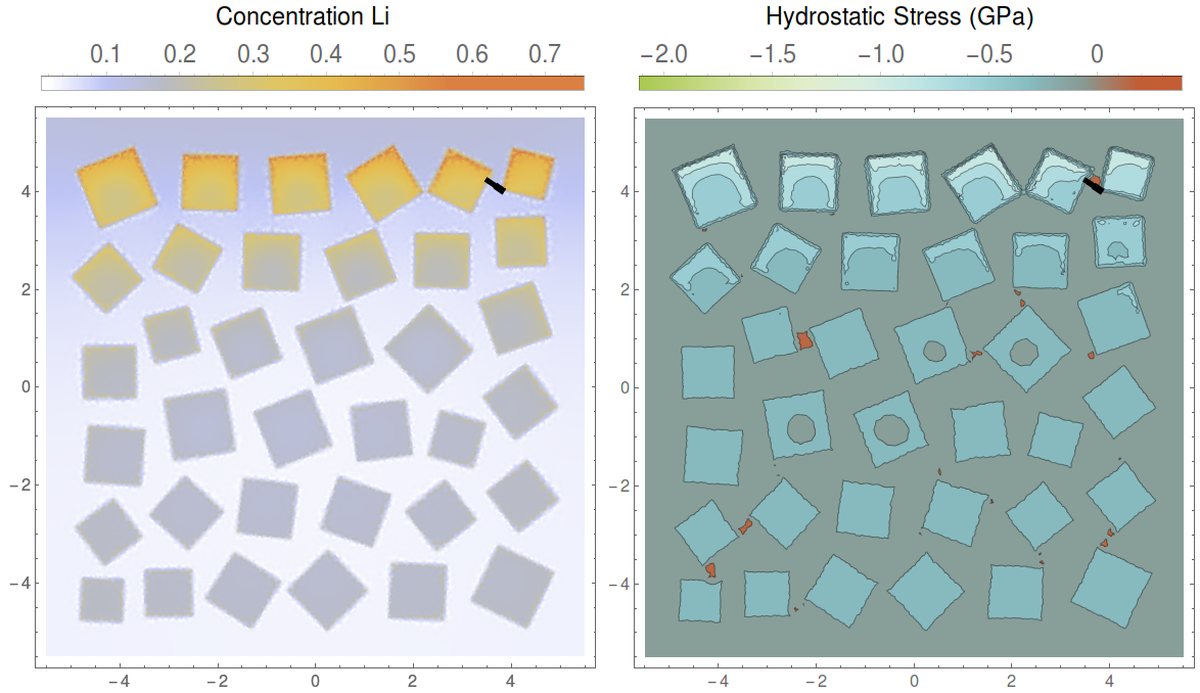} 
}

\subfloat[t = 400 s, SOC = 0.168185] {
\label{fig:fractureSE_2}
\includegraphics[width=0.45\textwidth]{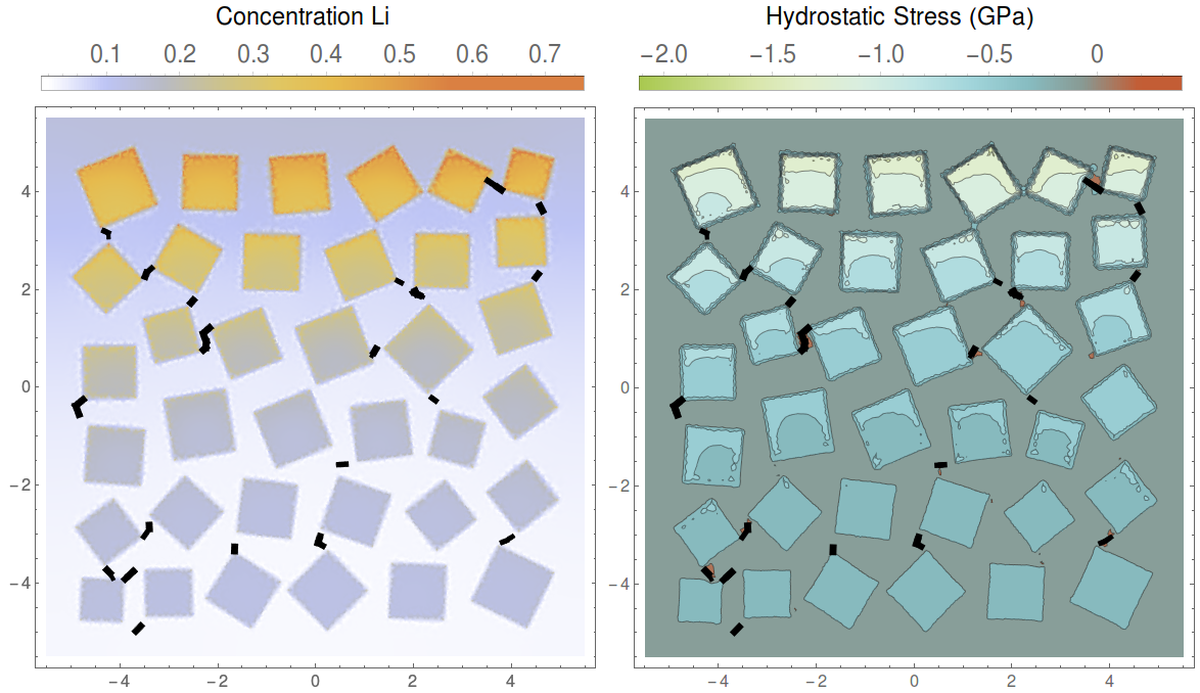} 
}

\subfloat[t = 900 s, SOC = 0.273068] {
\label{fig:fractureSE_3}
\includegraphics[width=0.45\textwidth]{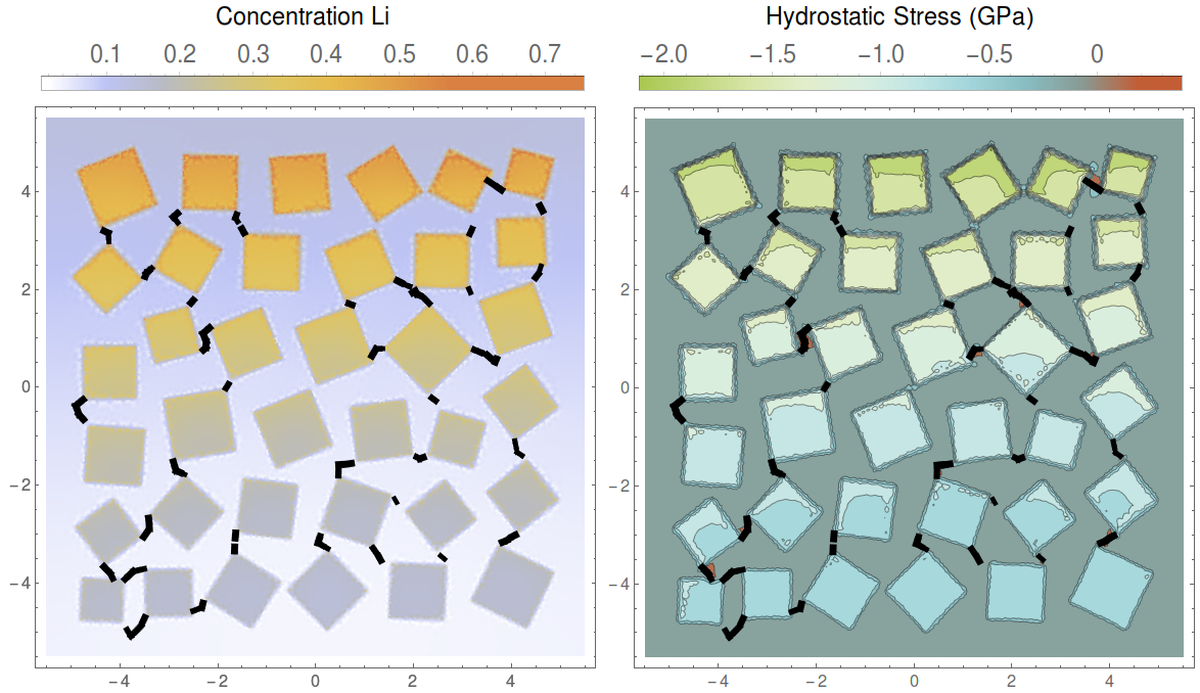} 
}
\caption{Evolution of damage in the solid electrolyte material at subsequent states of charge.
With increasing Li content, the active material undergoes chemical expansion. As the particles stress-free strain increases, compressive stress develops in most of the microstructure. However, a few regions in proximity of the particles corner are under tensile stress (brown regions in the right contour plots above). This tension grows large enough to initiate fracture in the solid electrolyte matrix, This phenomenon is captured in the simulation by the cohesive elements. Cracks (marked with black lines) propagate from corner to corner, cutting off diffusion paths for Li within the electrolyte.}
\label{fig:fractureSE_contours}
\end{figure}

A sequence of snapshots in Fig.~\ref{fig:fractureSE_contours} illustrate the state of charge (left column), the hydrostatic Cauchy stress (right column). 
Cracks propagating within the solid electrolyte material are represented as black lines. Thickening of the black lines represents progressive interface separation and accumulation of damage.

On average, compressive stress arises as a consequence of the particles' chemical expansion because the entire system is constrained by the surrounding material.
The simulation shows small regions of tensile stress developing in the area  near the particles' corners (rust-colored areas in the contour plots of Fig.~\ref{fig:fractureSE_contours}). The misalignment of these corners creates matrix shear- and tensile-stresses.  

Each snapshot in Fig.~\ref{fig:fractureSE_contours}  represents the lithiation and stress state at subsequent states of charge.
For all the numerical tests, the current density is held constant at $1$~mA~cm$^{-2}$ (a typical current density for commercial Li-ion batteries).
Time is indicative of state of charge, 
unless the evolution of stress varies significantly among tests. This a consequence of the electro-chemo-mechanical coupling~\cite{Bucci01012017}. 

As the state of charge progresses, the pressure in the particles increases. In Fig.~\ref{fig:fractureSE_3}, the compressive stress in the active material is higher than 1 GPa (for particles that have stored approximately 50\% of their total Li capacity).
The stress-strain curve measured by Sakuda \emph{et al.}~\cite{Sakuda2013youngMod}
shows a linear-elastic behavior  for Li$_2$S-P$_2$S$_5$ in compression. Our simulations predict the compressive stress
in the electrolyte 
to lie within the linear elastic range
of $0 - 200$ MPa, as measured by Sakuda \emph{et al.} 

\begin{table}[]
\caption{Material parameters for problems in Section~\ref{sec:fracture_SE}} 
\label{tab:fracture_SE}
 \resizebox{\columnwidth}{!}{
 \begin{tabular}{ll}
 \toprule
\textbf{Input value}	&	 \textbf{Description} \\
\midrule
$ F = 96485.3365$ C mol$^{-1}$  & Faraday's constant \\
$ R = 8.314$ J K$^{-1}$ mol$^{-1}$ & gas constant \\
$ T = 298$ K & temperature \\
$ M_{El} =  10^{-15}$ m$^2$ s$^{-1}$ & mobility of Li in the electrode material \\
$ M_{SE} =  10^{-13}$ m$^2$ s$^{-1}$ & mobility of Li in the solid electrolyte material \\
$ c_{max_{AM}} = 1 $ & maximum relative number of mole of Li per mole of electrode compound \\ 
$ c_{max_{SE}} = 0.25 $ & maximum relative number of mole of Li per mole of solid electrolyte \\ 
$ i = 10 $ A m$^2$ & maximum relative number of mole of Li per mole of solid electrolyte \\ 
%$ \rho_{H} = 2.6 \cdot 10^{4} $ mol m$^{-3} $ & molar density of the hosting material \\
$ \gamma_{Li} = 1$	& activity coefficient \\
$ \nu = 0.3 $ & Poisson's ratio for both materials \\
$ \beta_{AM} = 0.1 $ & relative lattice constant for Li in electrode material\\
$ \beta_{SE} = 0 $ & relative lattice constant for Li in the solid electrolyte material \\
$E_{AM} = 100$ GPa & Young's modulus of the electrode material\\
$E_{SE} = 15$ GPa & Young's modulus of the solid electrolyte material \\
$G_c = 1.0$ J m$^{-2}$ & fracture energy of the bulk solid electrolyte material \\
$\delta_0 = 5 $nm & opening displacement at the onset of damage \\
$\delta_{cr} = 20 \cdot \delta_0$  & critical opening displacement (complete interface separation) \\
 \bottomrule
\end{tabular}
}
\end{table}

We performed a series of numerical tests by varying the fracture energy of the SE material in the range $G_c = 0.25 - 4.0$ J m$^{-2}$.
%\gioshort{The extension of the damage predicted at various states of charge strongly depends on the SE fracture properties.} 
%Micro-cracking occurs in all the cases where electrode-particle expansion reaches 30\%.
%Fracture nucleation is delayed to higher lithiation values as the toughness of the solid electrolyte material increases.
%
To our knowledge, the only experimental data on the fracture toughness of a sulfide SE material is reported by McGrogan \emph{et al.}~\cite{McGrogan2016}. McGrogan and coauthors measured the toughness K$_{Ic}$ = $0.23 \pm 0.04$ MPa m$^{0.5}$ via nano-indentation of a glassy Li$_2$S-P$_2$S$_5$ sample. Such a fracture toughness corresponds to the fracture energy 
G$_c = 2.8 \pm 1.8 $ J m$^{-2}$, given the Young's modulus
E$_{SE}$ = $18.5 \pm 0.9$ GPa measured by McGrogan \emph{et al.}
%
%Highly conductive sulfide electrolytes react in contact with the atmosphere, making mechanical tests very challenging. 

%

\begin{figure}
\includegraphics[width=0.45\textwidth]{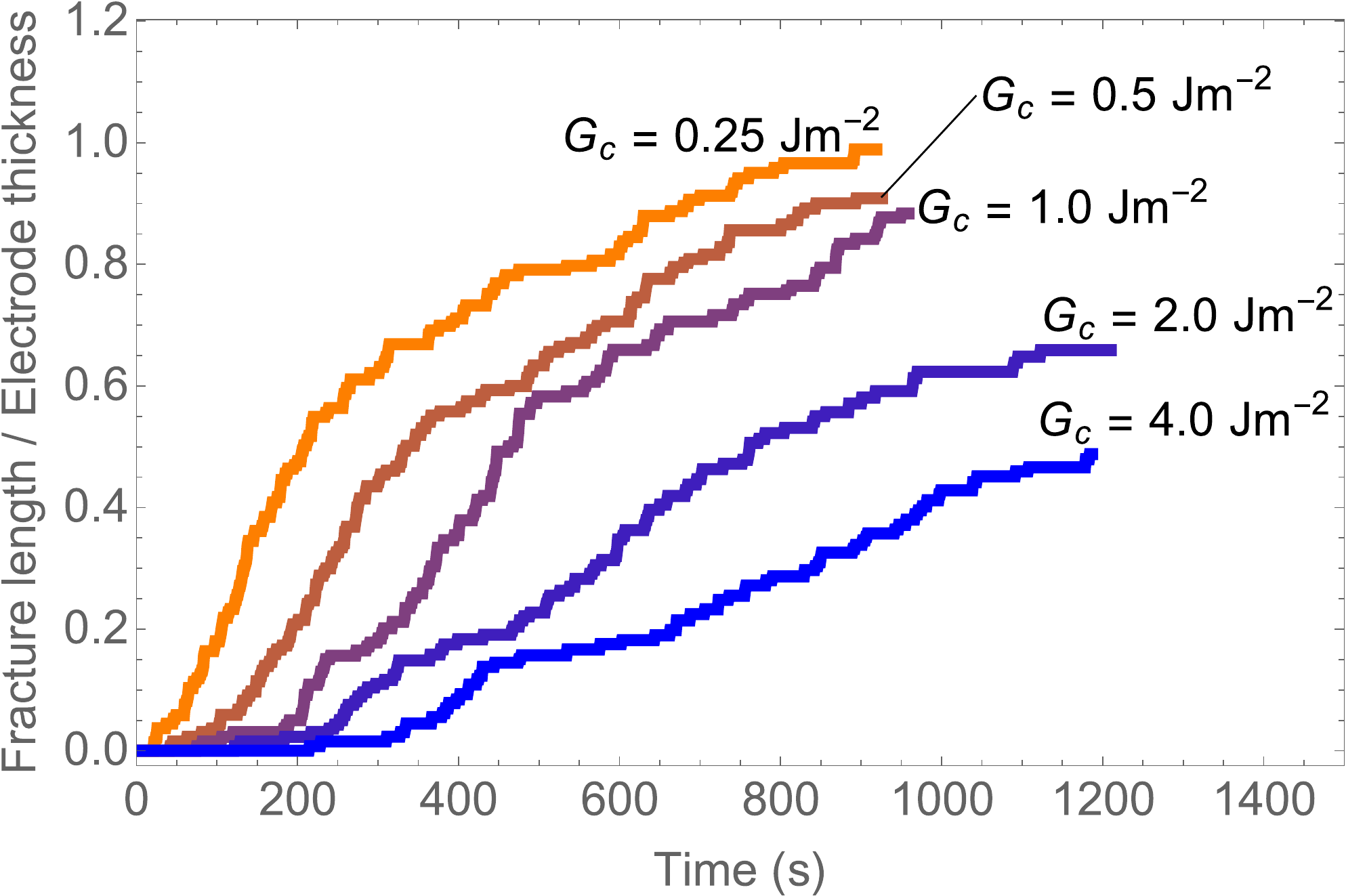} 
\caption{The extension of micro-cracks within the solid electrolyte material has been computed by a mesoscale finite element model. The curves represent five cases of SE fracture energies in the range $G_c = 0.25 - 4.0 $ J m$^{-2}$. 
%For each curve, it is possible to identify three stages: a)onset of fracture, b)approximately constant propagation rate, c) decreasing crack velocity up to saturation. As expected, crack nucleation is delayed in tougher materials. 
The propagation rate (slope of the curve in stage b) increases with decreasing fracture energy.}
\label{fig:crackL_vsFractureEnergy}
\end{figure}

The relative crack length (extension of fracture normalized with respect to electrode thickness) is illustrated in Fig.~\ref{fig:crackL_vsFractureEnergy} as it evolves with respect to time. 
For each curve in Fig.~\ref{fig:crackL_vsFractureEnergy} it is possible to identify three stages: \emph{a)}~onset of fracture, \emph{b)}~approximately constant propagation rate, and~\emph{c)}~decreasing propagation rate up to saturation. As expected, crack nucleation is delayed in tougher materials. The propagation rate (slope of the curve in stage \emph{b}) increases with decreasing fracture energy. 
In all the examples fracture propagates in a stable fashion (rather than sudden failure).
A plateau in the curves of Fig.~\ref{fig:crackL_vsFractureEnergy} 
indicates crack-growth saturation.  
In the cases with lower fracture energy, stage \emph{c} may be biased by the availability of crack patterns--even if about 10\% of the pre-inserted cohesive interfaces remain unfractured.
The pre-insertion of cohesive elements in specified locations is a possible shortcoming of this model.
As shown in Fig.~\ref{fig:crackL_vsFractureEnergy}, solid electrolytes with fracture energy up to $4$ J m$^{-2}$ are predicted to fracture when electrode particles undergo 30\% of intercalation-induced swelling .
%We refer to a future publication the analysis of the effect of SE mechanical degradation on the cell tortuosity and power density.
%Fracture will increase the tortuosity. The correlation between fracture and tortuosity is beyond the scope of this paper. However, we will discuss some of the implications of our findings in section~\ref{sec:totuosity_3d}.

\begin{figure}
\includegraphics[width=0.45\textwidth]{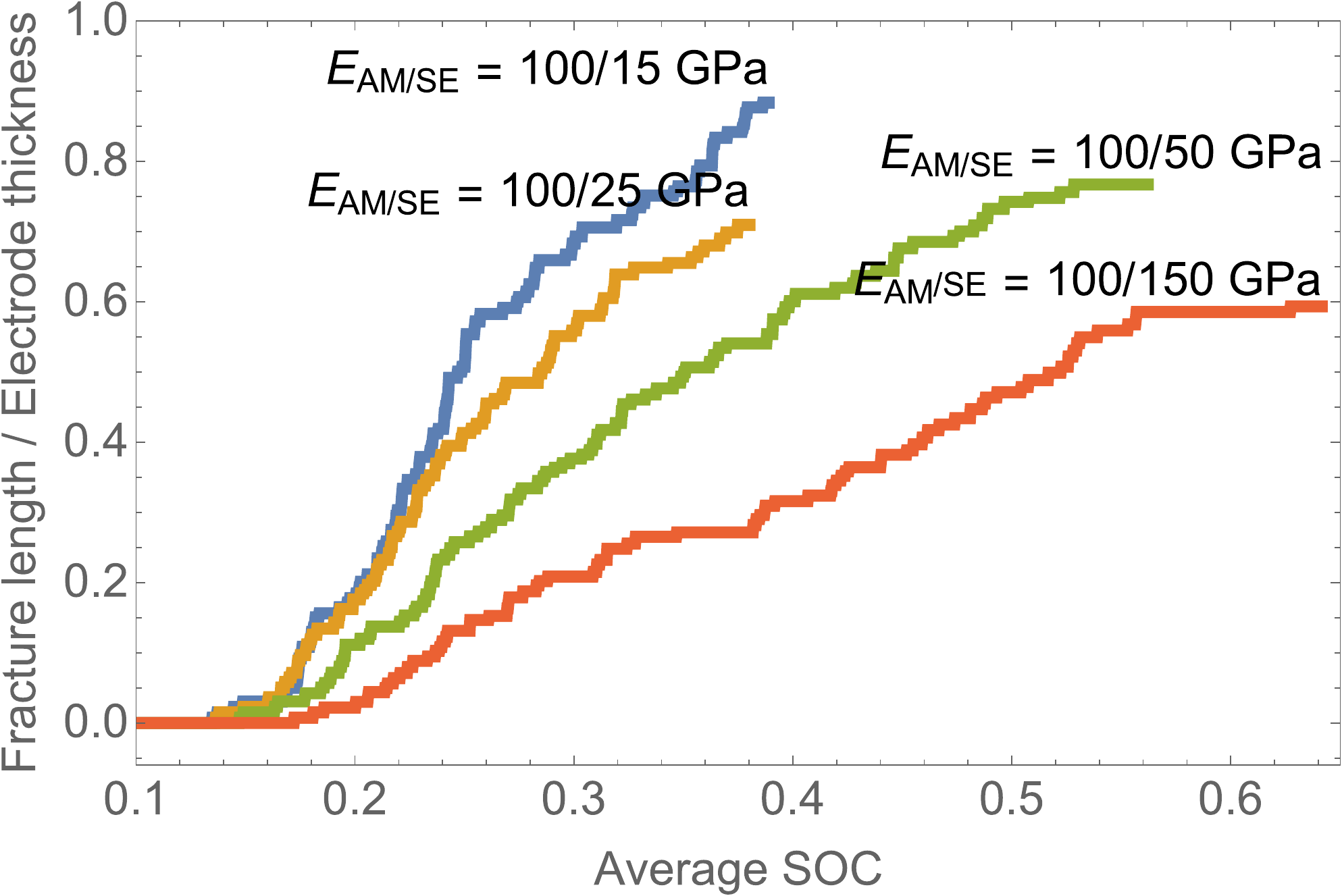} 
\caption{The extension of crack (normalized with respect to the electrode thickness) is plotted for three cases characterized by different Young's moduli of the solid electrolyte material, i.e., $E_{SE} = 15, 25, 50$ GPa. Other input parameters are the same of the baseline case (see Table~\ref{tab:fracture_SE}), and the Young's modulus of the active material remain fixed at $E_{SE} = 100$ GPa. As the stiffness of the SE increases, a lower velocity of fracture propagation is predicted. 
A stiffer SE tends to contain the chemical expansion of the active material. Shearing and tension --responsible for crack growth-- are less likely to arise in a less deformable material.}
\label{fig:crackL_vsElasticProp}
\end{figure}

We explore the dependence of the predicted damage on the electrolyte elastic properties. 
We consider electrolytes with 
Young's moduli $E_{SE} = 25$ GPa and $E_{SE} = 50$ GPa and $E_{SE} = 150$ GPa in addition to the baseline case ($E_{SE} = 15$ GPa).
The Young's modulus $E_{SE} = 150$ GPa is representative of a garnet solid electrolyte material (see Table~\ref{tab:SEmechanicalProp}).
%the electrode elastic properties by increasing the Young's modulus of the active material from $E_{AM} = 100 $ GPa to $E_{AM} = 200$ GPa. We also consider the case of a stiffer solid electrolyte: analyses with $E_{SE} = 25$ GPa  and $E_{SE} = 50$ GPa were carried out.
%
%
The results in Fig.~\ref{fig:crackL_vsElasticProp} illustrate the inverse relationship between the velocity of crack propagation and the electrolyte's stiffness. 
A more compliant solid electrolyte 
tends to deform more by stretching and shearing in response to the particles' volume change. 
%In addition to swelling, particles slide and rotate, because of their  random  morphological orientation and their non-uniform lithiation and chemical expansion. These large deformations are the source of the shear and tensile stresses and the reason that a non-linear model is required.
Regions of tensile-stress form in the SE matrix where fracture is promoted.  
An electrolyte with stiffness closer to that of the active material (here the Young's modulus of the electrode material is set to $E_{AM} = 100$ GPa) tends to develop higher compressive stress, but undergoes lower tension.
If the problem is considered from a linear elasticity perspective, these results may seem counter-intuitive.  However in the non-linear formulation, the large displacements give rise to tensile and shear stresses, particularly large in the case of compliant SEs.
 
In order to identify conditions that prevent mechanical degradation, we consider active materials with lower volume change associated with changes in Li stoichiometry. 
Furthermore, we raised the solid electrolytes fracture energy up to $G_c = 4.0$ J m$^{-2}$.
The model predicts that fracture is suppressed, for active materials with 7.5\% Vegard's expansion and solid electrolytes with the fracture energy $G_c = 4.0$ J m$^{-2}$ and Young's modulus $E_{SE} = 15$ GPa.
Thus, mechanical damage is predominantly dependent on SE fracture properties,
%The most restrictive condition is given by the SE fracture properties, 
as most Li-storage compounds 
have Vegard's expansion below 7.5\%~\cite{woodford:thesis,Mukhopadhyay2014}.

\begin{figure}
\includegraphics[width=0.45\textwidth]{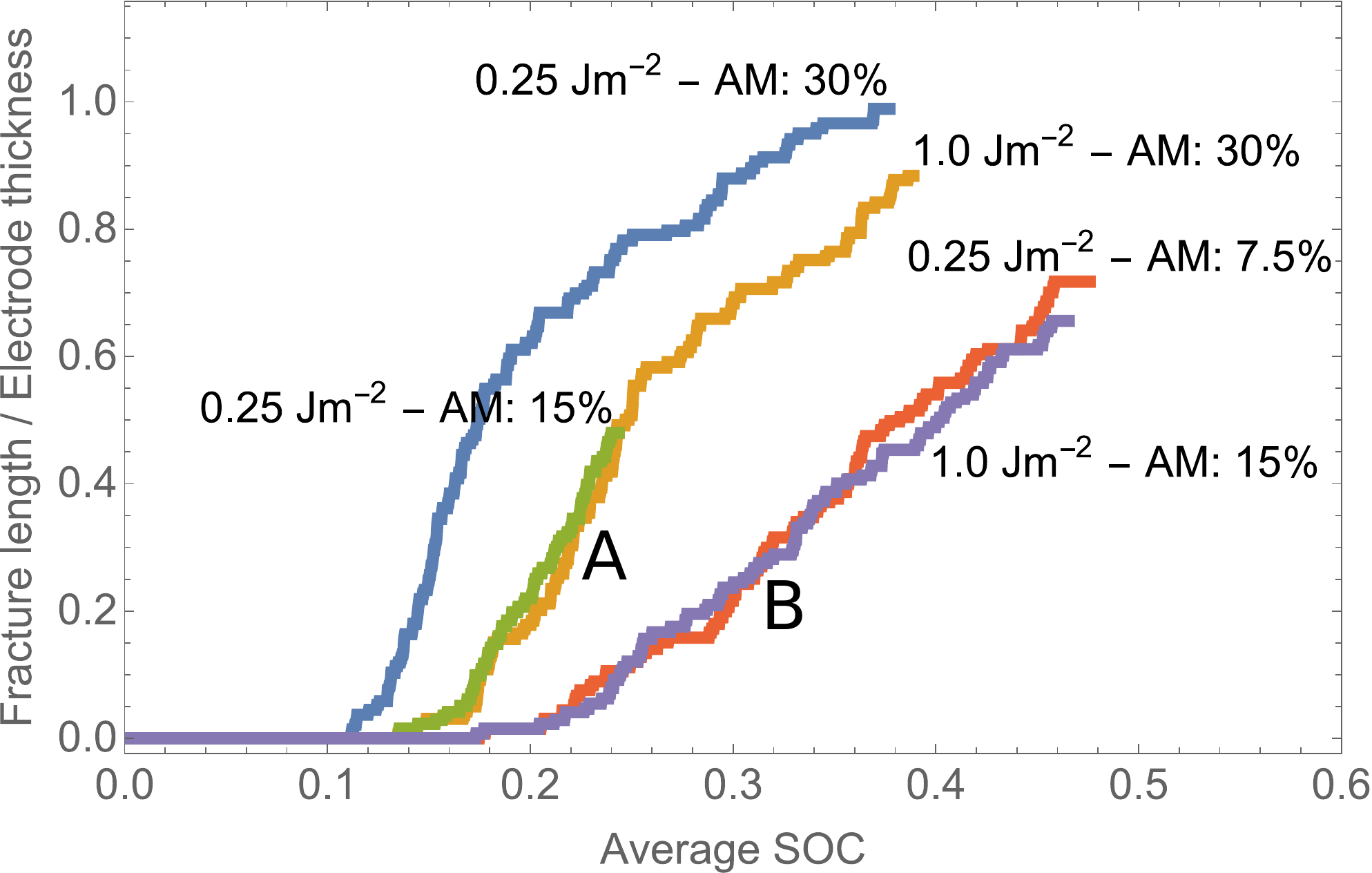} 
\captionsetup{singlelinecheck=off}
\caption{Various combination Vegard's parameters of the active material and SE fracture energy are analyzed. The overlap of two sets of results is observed.  
They correspond to the two cases marked as a
\newline 
\textbf{a)} 30 \% volume change of the active material and SE fracture energy $G_c = 1.0$ J m$^{-2}$ 
\newline 
\textbf{a)} 15 \% volume change of the active material and SE fracture energy $G_c = 0.25$ J m$^{-2}$ 
\newline
%\end{itemize}
and other two cases, marked as b 
\newline
\textbf{b)} 15 \% volume change of the active material and SE fracture energy $G_c = 1.0$ J m$^{-2}$ 
\newline
\textbf{b)} 7.5 \% volume change of the active material and SE fracture energy $G_c = 0.25$ J m$^{-2}$ 
\newline
The interpretation of this outcome is based on the system's energy balance.   
As the particles volume increases upon lithiation, the solid electrolyte volume is forced to shrink.
Here we assume the volume-averaged deformation of the entire region to be zero. A less restrictive assumption would allow for the electrode's thickness to evolve with state of charge.
In linear eleasticity, the hydrostatic pressure developing in both materials scales linearly with their volume change --proportionally to their bulk modulus.
It follows that the elastic energy stored in the SE material become four times larger, when the particles' expansion is doubled. If the fracture energy required to open new cracks is also four times larger,
fracture propagates at the same rate.
}
\label{fig:crackL_vsExpansion}
\end{figure}

Results from several tests are collected in Fig.~\ref{fig:crackL_vsExpansion}, where we observe the overlapping of two groups of curves.  
These curves are marked with a and b in Fig.~\ref{fig:crackL_vsExpansion}, and represent the tests with SE fracture energy $G_c = 1.0$ J m$^{-2}$ and $G_c = 0.25$ J m$^{-2}$.
The results can be explained by referring to the system's energy balance.
As the particle's volume increases upon lithiation, the solid electrolyte volume is forced to shrink because 
the electrode's volume-averaged deformation is zero.
%Linear elastic behavior and constant elastic properties (independent of Li content) are assumed for the electrode and electrolyte materials--this hypothesis for the SE is justified by measurements of  Li$_2$S-P$_2$S$_5$ stress-strain curve~\cite{Sakuda2013youngMod}.
%
Hydrostatic pressure developing in the electrode and electrolyte materials scales linearly with volume change. The pressure is also proportional to the materials' bulk moduli.
It follows that the elastic energy stored in the electrolyte becomes four times larger when the active material's expansion is doubled. If the fracture energy required to open new cracks is also four times larger,
fracture propagation rate remains the same. 
In order to characterize the dependence on the total energy, both stored elastic and fracture energy, we define a new dimensionless parameter for all-solid-state electrodes: 
$ \mathcal{G} = 0.5 k_{SE} (3 \beta_{AM} A_{AM})^2 / (H G_c)$.
%
%In other words the overlapping cases \emph{a} and \emph{b} are characterized by the same value of the dimensionless parameter
%$ \mathcal{G} = 0.5 k_{SE} (3 \beta_{AM} A_{AM})^2 / (H G_c)$,
%
In the definition of $ \mathcal{G}$, $k_{SE}$ is the SE bulk modulus, $ \beta_{AM} $ the Vegard's parameter ($3 \beta_{AM}$ is the volumetric expansion rate) of the active material, $H$ is the electrode's thickness, and $A_{AM}$ is the area of active material. 
The two overlapping curves for the case \emph{a} and the
the case \emph{b} in  Fig.~\ref{fig:crackL_vsExpansion} are characterized by the same value of $\mathcal{G}$. Therefore, the dimensionless parameter $\mathcal{G}$ can be used to generalize the results presented here.
%The fact that the model is able to capture this relationship for both cases \emph{a} and \emph{b} presented in Fig.~\ref{fig:crackL_vsExpansion} is a demonstration of its consistency.

%%%%%%%%%%%%%%%%%%%%%%%%%%%%%%%%%%%%%%%%%%%

\section{Conclusions}
\label{Conclusions}

%We employed a coupled electro-chemo-mechanical model to quantify the material properties and usage conditions that cause a solid state electrolyte to fracture.
%
%Our coupled electro-chemo-mechanical model predicts that solid-state composite  electrodes will fracture during electrochemical cycling.
Electro-chemo-mechanical FEM simulations capture the onset and propagation of damage in a solid-state composite electrode.
Fracture is prevented if electrode-particle's expansion is lower than 7.5\% and the solid-electrolyte's fracture energy higher than $G_c = 4$ J m$^{-2}$ (under the assumption of SE Young's modulus E$_{SE} = 15$ GPa). 
This condition restricts the choice of electrolyte based on its fracture properties, 
while most intercalation oxides  
have volume expansion below 7.5\%~\cite{woodford:thesis}. 
We refer here to the average volume change of a poly-crystalline material--Vegard's parameters can be largely anisotropic, this is for instance the case of graphite. 

Here we refer to the aver
Intercalation-induced expansion of the active material is constrained in dense solid-state electrodes and  electrolytes are prone to
mechanical degradation. 
The particles and SE together create a microstructure--the shape of particles and their proximity within the microstructure determine fracture.
Microstructural inhomogeneities, such as particle-to-particle misalignment and particle asperities are sufficient to cause tensile and shear stress in the solid electrolyte matrix.

The simulations predict fracture to propagate in a stable fashion (rather than abruptly).
As expected, crack nucleation is delayed in tougher materials. The propagation rate and the final extension of cracks also decrease with increasing electrolyte fracture energy. 

Perhaps counter-intuitively, the analyses show that 
compliant solid electrolytes (with Young's modulus in the order of E$_{SE} = 15$ GPa) are more prone to micro-cracking. 
Shearing and stress arise 
if the particles have surface asperities
or the stress-fields of nearby particles interact.
% and they undergo non-uniform lithiation and chemical expansion. 
Because compliant SEs allow for large deformations
%lithiation causes large deformations
%stresses cause the particles to translate and rotate,
they are more likely to develop localized tension and  fracture.
A non-linear kinematics model is required to predict this effect
%This effect can be captured by formulating the problem in nonlinear kinematics, and it contradicts 
which contradicts the speculation that sulfide SEs are more suitable
for the design of bulk-type batteries than oxide SEs~\cite{Sakuda2013}.

To our knowledge, this work is the first to investigate mechanical reliability of all-solid state batteries. 
The results presented have implications for the battery power-density. Fracture in solid Li-ion conductors represents a barrier for Li transport and accelerate the decay of rate performance.
%By physically occluding the dendrite path, solid electrolytes combined with a Li-metal anode offer an opportunity to 
%prevent destruction of the cell by shorting.
%However, microstructures need to be produced with controlled distribution of flaws ad pores.

%Future developments of this work include a 3D analysis of microstructures based on X-ray tomography data, and a detailed study on the effect of micro-cracks on the electrode's effective diffusivity.

Reliability of ASSBs will depend on
the elastic energy associated with intercalation-induced strain, the solid-electrolyte fracture energy and the geometry of the microstructure.
Therefore, a simple design rule can be based on the dimensionless parameter 
$ \mathcal{G} = 0.5 k_{SE} (3 \beta_{AM} A_{AM})^2 / (H G_c) $ representing the ratio between elastic and fracture energies.
We predict the integrity of elastic-brittle solid-state electrolytes to be preserved 
when the condition
 $ \mathcal{G} < 1000$ is met.

%%%%%%%%%%%%%%%%%%%%%%%%%%%%%%%
\section*{Acknowledgments}
The work was supported by the grant DE-SC0002633 funded by the U.S. Department of Energy, Office of Science.

%%%REFERENCES%%%

%\bibliography{All-solid_batteries_biblio} %You need to replace "rsc" on this line with the name of your .bib file

\begin{thebibliography}{10}

\bibitem{Takada:2013}
K.~Takada, ``Progress and prospective of solid-state lithium batteries,'' {\em
  Acta Materialia}, vol.~61, no.~3, pp.~759 -- 770, 2013.

\bibitem{Li2015}
J.~Li, C.~Ma, M.~Chi, C.~Liang, and N.~J. Dudney, ``{Solid electrolyte: The key
  for high-voltage lithium batteries},'' {\em Advanced Energy Materials},
  vol.~5, no.~4, pp.~1--6, 2015.

\bibitem{Kim2015}
J.~G. Kim, B.~Son, S.~Mukherjee, N.~Schuppert, A.~Bates, O.~Kwon, M.~J. Choi,
  H.~Y. Chung, and S.~Park, ``{A review of lithium and non-lithium based solid
  state batteries},'' {\em Journal of Power Sources}, vol.~282, pp.~299--322,
  2015.

\bibitem{ASSBgarnet_rev2014}
V.~Thangadurai, S.~Narayanan, and D.~Pinzaru, ``Garnet-type solid-state fast li
  ion conductors for li batteries: critical review,'' {\em Chem. Soc. Rev.},
  vol.~43, pp.~4714--4727, 2014.

\bibitem{C0CS00081G}
E.~Quartarone and P.~Mustarelli, ``Electrolytes for solid-state lithium
  rechargeable batteries: recent advances and perspectives,'' {\em Chem. Soc.
  Rev.}, vol.~40, pp.~2525--2540, 2011.

\bibitem{Takada2014}
K.~Takada, N.~Ohta, and Y.~Tateyama, ``{Recent Progress in Interfacial
  Nanoarchitectonics in Solid-State Batteries},'' {\em Journal of Inorganic and
  Organometallic Polymers and Materials}, pp.~205--213, 2014.

\bibitem{Tatsumisago2013}
M.~Tatsumisago, M.~Nagao, and A.~Hayashi, ``{Recent development of sulfide
  solid electrolytes and interfacial modification for all-solid-state
  rechargeable lithium batteries},'' {\em Journal of Asian Ceramic Societies},
  vol.~1, pp.~17--25, mar 2013.

\bibitem{Suzuki2015172}
Y.~Suzuki, K.~Kami, K.~Watanabe, A.~Watanabe, N.~Saito, T.~Ohnishi, K.~Takada,
  R.~Sudo, and N.~Imanishi, ``Transparent cubic garnet-type solid electrolyte
  of al2o3-doped li7la3zr2o12,'' {\em Solid State Ionics}, vol.~278, pp.~172 --
  176, 2015.

\bibitem{bucciActaMat2015}
G.~Bucci, Y.-M. Chiang, and W.~Carter, ``Formulation of the coupled
  electrochemical-mechanical boundary-value problem, with applications to
  transport of multiple charged species,'' {\em Acta Materialia}, vol.~62,
  pp.~33--51, 2016.

\bibitem{bucci2016book}
G.~Bucci and W.~Carter, {\em Mechanics of Materials. Micro-mechanics in
  electrochemical systems (in production)}.
\newblock Springer, 2016.

\bibitem{Herbert2011}
E.~G. Herbert, W.~E. Tenhaeff, N.~J. Dudney, and G.~M. Pharr, ``{Mechanical
  characterization of LiPON films using nanoindentation},'' {\em Thin Solid
  Films}, vol.~520, no.~1, pp.~413--418, 2011.

\bibitem{Cho2012}
Y.-H. Cho, J.~Wolfenstine, E.~Rangasamy, H.~Kim, H.~Choe, and J.~Sakamoto,
  ``{Mechanical properties of the solid Li-ion conducting electrolyte:
  Li$_{0.33}$La$_{0.57}$TiO$_3$},'' {\em Journal of Materials Science},
  vol.~47, pp.~5970--5977, 2012.

\bibitem{Ni2012}
J.~E. Ni, E.~D. Case, J.~S. Sakamoto, E.~Rangasamy, and J.~B. Wolfenstine,
  ``{Room temperature elastic moduli and Vickers hardness of hot-pressed LLZO
  cubic garnet},'' {\em Journal of Materials Science}, vol.~47, pp.~7978--7985,
  2012.

\bibitem{Sakuda2013youngMod}
A.~Sakuda, A.~Hayashi, and Y.~Takigawa, ``{Evaluation of elastic modulus of
  Li$_2$S - P$_2$S$_5$ glassy solid electrolyte by ultrasonic sound velocity
  measurement and compression test},'' {\em Journal of the Ceramic Society of
  Japan}, vol.~121, no.~11, pp.~946--949, 2013.

\bibitem{McGrogan2016}
F.~P. McGrogan, T.~Swamy, S.~R. Bishop, E.~Eggleton, L.~Porz, X.~Chen, Y.-M.
  Chiang, and K.~J.~V. Vliet, ``Compliant yet brittle mechanical behavior of
  {Li$_2$S-P$_2$S$_5$} lithium-ion conducting solid electrolyte,'' {\em
  Advanced Energy Materials}, p.~1602011, 2017.

\bibitem{Wang2014}
Z.~Q. Wang, M.~S. Wu, G.~Liu, X.~L. Lei, B.~Xu, and C.~Y. Ouyang, ``{Elastic
  properties of new solid state electrolyte material Li$_10$GeP$_2$S$_12$: A
  study from first-principles calculations},'' {\em International Journal of
  Electrochemical Science}, vol.~9, no.~2, pp.~562--568, 2014.

\bibitem{woodford2010electrochemical}
W.~H. Woodford, Y.-M. Chiang, and W.~C. Carter, ``Electrochemical shock of
  intercalation electrodes: a fracture mechanics analysis,'' {\em Journal of
  The Electrochemical Society}, vol.~157, no.~10, pp.~A1052--A1059, 2010.

\bibitem{Bhandakkar20101424}
T.~K. Bhandakkar and H.~Gao, ``Cohesive modeling of crack nucleation under
  diffusion induced stresses in a thin strip: Implications on the critical size
  for flaw tolerant battery electrodes,'' {\em International Journal of Solids
  and Structures}, vol.~47, no.~10, pp.~1424 -- 1434, 2010.

\bibitem{BowerGuduru:2012}
A.~F. Bower and P.~R. Guduru, ``A simple finite element model of diffusion,
  finite deformation, plasticity and fracture in lithium ion insertion
  electrode materials,'' {\em Modelling and Simulation in Materials Science and
  Engineering}, vol.~20, no.~4, p.~045004, 2012.

\bibitem{Renganathan01022010}
S.~Renganathan, G.~Sikha, S.~Santhanagopalan, and R.~E. White, ``Theoretical
  analysis of stresses in a lithium ion cell,'' {\em Journal of The
  Electrochemical Society}, vol.~157, no.~2, pp.~A155--A163, 2010.

\bibitem{Garcia01012005}
R.~E. Garc\'{i}a, Y.-M. Chiang, W.~Craig~Carter, P.~Limthongkul, and C.~M.
  Bishop, ``Microstructural modeling and design of rechargeable lithium-ion
  batteries,'' {\em Journal of The Electrochemical Society}, vol.~152, no.~1,
  pp.~A255--A263, 2005.

\bibitem{Wang01112007}
C.-W. Wang and A.~M. Sastry, ``Mesoscale modeling of a li-ion polymer cell,''
  {\em Journal of The Electrochemical Society}, vol.~154, no.~11,
  pp.~A1035--A1047, 2007.

\bibitem{Golmon20091567}
S.~Golmon, K.~Maute, and M.~L. Dunn, ``Numerical modeling of
  electrochemical-mechanical interactions in lithium polymer batteries,'' {\em
  Computers \& Structures}, vol.~87, no.~23-24, pp.~1567 -- 1579, 2009.

\bibitem{Zhu01012012}
M.~Zhu, J.~Park, and A.~M. Sastry, ``Fracture analysis of the cathode in li-ion
  batteries: A simulation study,'' {\em Journal of The Electrochemical
  Society}, vol.~159, no.~4, pp.~A492--A498, 2012.

\bibitem{Purkayastha2012}
R.~T. Purkayastha and R.~M. McMeeking, ``An integrated 2-d model of a lithium
  ion battery: the effect of material parameters and morphology on storage
  particle stress,'' {\em Computational Mechanics}, vol.~50, no.~2,
  pp.~209--227, 2012.

\bibitem{NME:NME5133}
C.~Miehe, H.~Dal, L.-M. Sch\"{a}nzel, and A.~Raina, ``A phase-field model for
  chemo-mechanical induced fracture in lithium-ion battery electrode
  particles,'' {\em International Journal for Numerical Methods in
  Engineering}, pp.~n/a--n/a, 2015.
\newblock nme.5133.

\bibitem{Aifantis2005203}
K.~Aifantis and J.~Dempsey, ``Stable crack growth in nanostructured
  li-batteries,'' {\em Journal of Power Sources}, vol.~143, no.~12, pp.~203 --
  211, 2005.

\bibitem{Ryu20111717}
I.~Ryu, J.~W. Choi, Y.~Cui, and W.~D. Nix, ``Size-dependent fracture of si
  nanowire battery anodes,'' {\em Journal of the Mechanics and Physics of
  Solids}, vol.~59, no.~9, pp.~1717 -- 1730, 2011.

\bibitem{Aifantis20112122}
K.~Aifantis and S.~Hackney, ``Mechanical stability for nanostructured sn- and
  si-based anodes,'' {\em Journal of Power Sources}, vol.~196, no.~4, pp.~2122
  -- 2127, 2011.

\bibitem{Kalnaus20118116}
S.~Kalnaus, K.~Rhodes, and C.~Daniel, ``A study of lithium ion intercalation
  induced fracture of silicon particles used as anode material in li-ion
  battery,'' {\em Journal of Power Sources}, vol.~196, no.~19, pp.~8116 --
  8124, 2011.

\bibitem{Xia201478}
Y.~Xia, T.~Wierzbicki, E.~Sahraei, and X.~Zhang, ``Damage of cells and battery
  packs due to ground impact,'' {\em Journal of Power Sources}, vol.~267,
  pp.~78 -- 97, 2014.

\bibitem{Zhang201547}
X.~Zhang and T.~Wierzbicki, ``Characterization of plasticity and fracture of
  shell casing of lithium-ion cylindrical battery,'' {\em Journal of Power
  Sources}, vol.~280, pp.~47 -- 56, 2015.

\bibitem{Hao2013415}
F.~Hao and D.~Fang, ``Reducing diffusion-induced stresses of
  electrode-collector bilayer in lithium-ion battery by pre-strain,'' {\em
  Journal of Power Sources}, vol.~242, pp.~415 -- 420, 2013.

\bibitem{Chew20144176}
H.~B. Chew, B.~Hou, X.~Wang, and S.~Xia, ``Cracking mechanisms in lithiated
  silicon thin film electrodes,'' {\em International Journal of Solids and
  Structures}, vol.~51, no.~23-24, pp.~4176 -- 4187, 2014.

\bibitem{Drozdov201467}
A.~Drozdov, ``Constitutive equations for self-limiting lithiation of electrode
  nanoparticles in li-ion batteries,'' {\em Mechanics Research Communications},
  vol.~57, pp.~67 -- 73, 2014.

\bibitem{Ye2014447}
J.~Ye, Y.~An, T.~Heo, M.~Biener, R.~Nikolic, M.~Tang, H.~Jiang, and Y.~Wang,
  ``Enhanced lithiation and fracture behavior of silicon mesoscale pillars via
  atomic layer coatings and geometry design,'' {\em Journal of Power Sources},
  vol.~248, pp.~447 -- 456, 2014.

\bibitem{Greve2012377}
L.~Greve and C.~Fehrenbach, ``Mechanical testing and macro-mechanical finite
  element simulation of the deformation, fracture, and short circuit initiation
  of cylindrical lithium ion battery cells,'' {\em Journal of Power Sources},
  vol.~214, pp.~377 -- 385, 2012.

\bibitem{Wang2012236}
Y.~Wang, Y.~He, R.~Xiao, H.~Li, K.~Aifantis, and X.~Huang, ``Investigation of
  crack patterns and cyclic performance of ti-si nanocomposite thin film anodes
  for lithium ion batteries,'' {\em Journal of Power Sources}, vol.~202,
  pp.~236 -- 245, 2012.

\bibitem{Pharr2014569}
M.~Pharr, Z.~Suo, and J.~J. Vlassak, ``Variation of stress with charging rate
  due to strain-rate sensitivity of silicon electrodes of li-ion batteries,''
  {\em Journal of Power Sources}, vol.~270, pp.~569 -- 575, 2014.

\bibitem{Min2015835}
J.~K. Min, M.~Stackpool, C.~H. Shin, and C.-H. Lee, ``Cell safety analysis of a
  molten sodium-sulfur battery under failure mode from a fracture in the solid
  electrolyte,'' {\em Journal of Power Sources}, vol.~293, pp.~835 -- 845,
  2015.

\bibitem{Zhang2015102}
C.~Zhang, S.~Santhanagopalan, M.~A. Sprague, and A.~A. Pesaran, ``Coupled
  mechanical-electrical-thermal modeling for short-circuit prediction in a
  lithium-ion cell under mechanical abuse,'' {\em Journal of Power Sources},
  vol.~290, pp.~102 -- 113, 2015.

\bibitem{Ma2015114}
Z.~Ma, Z.~Xie, Y.~Wang, P.~Zhang, Y.~Pan, Y.~Zhou, and C.~Lu, ``Failure modes
  of hollow core-shell structural active materials during the
  lithiation-delithiation process,'' {\em Journal of Power Sources}, vol.~290,
  pp.~114 -- 122, 2015.

\bibitem{Zhang2015309}
C.~Zhang, S.~Santhanagopalan, M.~A. Sprague, and A.~A. Pesaran, ``A
  representative-sandwich model for simultaneously coupled
  mechanical-electrical-thermal simulation of a lithium-ion cell under
  quasi-static indentation tests,'' {\em Journal of Power Sources}, vol.~298,
  pp.~309 -- 321, 2015.

\bibitem{Damle2016373}
S.~S. Damle, S.~Pal, P.~N. Kumta, and S.~Maiti, ``Effect of silicon
  configurations on the mechanical integrity of silicon-carbon nanotube
  heterostructured anode for lithium ion battery: A computational study,'' {\em
  Journal of Power Sources}, vol.~304, pp.~373 -- 383, 2016.

\bibitem{Laresgoiti}
I.~Laresgoiti, S.~Kabitz, M.~Ecker, and D.~U. Sauer, ``Modeling mechanical
  degradation in lithium ion batteries during cycling: Solid electrolyte
  interphase fracture,'' {\em Journal of Power Sources}, vol.~300, pp.~112 --
  122, 2015.

\bibitem{Ryu2014274}
I.~Ryu, S.~W. Lee, H.~Gao, Y.~Cui, and W.~D. Nix, ``Microscopic model for
  fracture of crystalline si nanopillars during lithiation,'' {\em Journal of
  Power Sources}, vol.~255, pp.~274 -- 282, 2014.

\bibitem{Dai}
Y.~Dai, L.~Cai, and R.~E. White, ``Simulation and analysis of stress in a
  li-ion battery with a blended limn2o4 and lini0.8co0.15al0.05o2 cathode,''
  {\em Journal of Power Sources}, vol.~247, pp.~365 -- 376, 2014.

\bibitem{Yang}
H.~Yang, F.~Fan, W.~Liang, X.~Guo, T.~Zhu, and S.~Zhang, ``A chemo-mechanical
  model of lithiation in silicon,'' {\em Journal of the Mechanics and Physics
  of Solids}, vol.~70, pp.~349 -- 361, 2014.

\bibitem{Dimitrijevic}
B.~Dimitrijevic, K.~Aifantis, and K.~Hackl, ``The influence of particle size
  and spacing on the fragmentation of nanocomposite anodes for li batteries,''
  {\em Journal of Power Sources}, vol.~206, pp.~343 -- 348, 2012.

\bibitem{Lee201637}
S.~Lee, J.~Yang, and W.~Lu, ``Debonding at the interface between active
  particles and \{PVDF\} binder in li-ion batteries,'' {\em Extreme Mechanics
  Letters}, vol.~6, pp.~37 -- 44, 2016.

\bibitem{Sakuda2013}
A.~Sakuda, A.~Hayashi, and M.~Tatsumisago, ``{Sulfide Solid Electrolyte with
  Favorable Mechanical Property for All-Solid-State Lithium Battery},'' {\em
  Scientific reports}, vol.~3, p.~2261, jan 2013.

\bibitem{C5TA08574H}
Y.~Zhu, X.~He, and Y.~Mo, ``First principles study on electrochemical and
  chemical stability of solid electrolyte-electrode interfaces in
  all-solid-state li-ion batteries,'' {\em J. Mater. Chem. A}, vol.~4,
  pp.~3253--3266, 2016.

\bibitem{RichardsCeder}
W.~D. Richards, L.~J. Miara, Y.~Wang, J.~C. Kim, and G.~Ceder, ``Interface
  stability in solid-state batteries,'' {\em Chemistry of Materials}, vol.~28,
  no.~1, pp.~266--273, 2016.

\bibitem{Wenzel201624}
S.~Wenzel, D.~A. Weber, T.~Leichtweiss, M.~R. Busche, J.~Sann, and J.~Janek,
  ``Interphase formation and degradation of charge transfer kinetics between a
  lithium metal anode and highly crystalline {Li$_7$P$_3$S$_11$} solid
  electrolyte,'' {\em Solid State Ionics}, vol.~286, pp.~24 -- 33, 2016.

\bibitem{woodford:thesis}
W.~H. Woodford, {\em Electrochemical Shock: Mechanical Degradation of
  Ion-Intercalation Materials}.
\newblock PhD thesis, Massachusetts Institute of Technology, 2013.

\bibitem{Mukhopadhyay2014}
A.~Mukhopadhyay and B.~W. Sheldon, ``{Deformation and stress in electrode
  materials for Li-ion batteries},'' {\em Progress in Materials Science},
  vol.~63, no.~February, pp.~58--116, 2014.

\bibitem{BangerthHartmannKanschat2007}
W.~Bangerth, R.~Hartmann, and G.~Kanschat, ``{deal.II} -- a general purpose
  object oriented finite element library,'' {\em ACM Trans. Math. Softw.},
  vol.~33, no.~4, pp.~24/1--24/27, 2007.

\bibitem{dealII82}
W.~Bangerth, T.~Heister, L.~Heltai, G.~Kanschat, M.~Kronbichler, M.~Maier,
  B.~Turcksin, and T.~D. Young, ``The \text{deal.II} library, version 8.2,''
  {\em Archive of Numerical Software}, vol.~3, 2015.

\bibitem{Harris2013}
S.~J. Harris and P.~Lu, ``{Effects of Inhomogeneities --Nanoscale to
  Mesoscale-- on the Durability of Li-Ion Batteries},'' {\em The Journal of
  Physical Chemistry C}, vol.~117, no.~13, pp.~6481--6492, 2013.

\bibitem{Bucci01012017}
G.~Bucci, T.~Swamy, S.~Bishop, B.~W. Sheldon, Y.-M. Chiang, and W.~C. Carter,
  ``The effect of stress on battery-electrode capacity,'' {\em Journal of The
  Electrochemical Society}, vol.~164, no.~4, pp.~A645--A654, 2017.

\end{thebibliography}
\bibliographystyle{ieeetr} %the RSC's .bst file

\end{document}